\pdfoutput=1
\documentclass[twocolumn,showpacs,floatfix,prl]{revtex4}

\usepackage[final]{graphicx}
\DeclareGraphicsExtensions{.pdf}
\usepackage{amsmath,amssymb,bm}
\usepackage{float}

\begin{document}

\title{Quantum spin Hall effect induced by non-magnetic and magnetic staggered potentials}

\author{Huaiming Guo$^{1*}$, Shiping Feng$^2$ and Shun-Qing Shen$^3$}
\affiliation{$^1$Department of Physics and Center of Theoretical Physics, Capital Normal University,
Beijing, 100048, China}
\affiliation{$^2$Department of Physics, Beijing Normal University,
Beijing, 100875, China}
\affiliation{$^3$Department of Physics and Center of Computational and Theoretical Physics, The University of Hong Kong, Pokfulam Road, Hong Kong}

\begin{abstract}
We have a comparative study of the quantum spin Hall (QSH) effects
induced by non-magnetic and magnetic staggered potentials
respectively and show that they have the same effect in driving the
topological phase transition. The result implies that both time-reversal (${\cal T}$) preserving and breaking systems can host QSH effect.
We also investigate the stability of the resulting QSH effect to disorder and find that for ${\cal T}$ invariant system the edge states are
always robust while those of ${\cal T}$ breaking system are also robust if there is additional symmetry in the system.
\end{abstract}

\pacs{73.43.-f, 72.25.Hg, 73.20.-r, 85.75.-d}
\maketitle

Recently the field of topological insulator (TI) has attracted a
great deal of interests, due to their exotic physical properties as
well as potential applications, such as spintronics, quantum
computing, et al \cite{moore1, hasan1, xlqi1}. Many materials have
been predicted and discovered to show TI phases (including
$HgTe/CdTe$ quantum wells (QWs) \cite{bernevig1, konig1}, bismuth
antimony alloys \cite{hsieh1, hsieh2}; $Bi_2Se_3$, $Bi_2Te_3$, and
$Sb_2Te_3$ \cite{hjzhang1, yxia1, ylchen1, hsieh3}; Heusler
compounds \cite{hlin1, chadov1}; Tl-based ternary chalcogenide
series \cite{hlin2, bhyan1}; et al). The findings of the real
materials not only provide a platform to test the predictions of
many unusual phenomena exhibited by TI \cite{lfu1, xlqi2, essin1,
mfranz1}, but also inspire more theoretical studies on TI.

The study of TI begins at the theoretical proposal of QSH effect in
graphene by Kane and Mele \cite{kane1, kane2}. They expected that
spin-orbit coupling will convert graphene from an ideal 2D
semimetallic state to a QSH insulator. The resulting QSH insulator
is topologically distinct from a band insulator, so it is referred
as TI. However the calculations have suggested that the spin-orbit
coupling in graphene is too small to reveal the QSH effect
experimentally \cite{ygyao1}. Remarkably, in 2007 the QSH effect was
realized in $HgTe/CdTe$ QWs following the theoretical suggestion of
Bernevig, Hughes and Zhang \cite{bernevig1, konig1}. Later though
many studies have been carried out in identifying new physical
systems that will possess topological nontrivial phases, till now
the 2D TI is only found experimentally in $HgTe/CdTe$ QWs. In the
low-energy effective theory, the QSH effect can be understood from
Dirac Hamiltonian with masses \cite{wyshan1, sqshen1}. The relative
signs of the masses at the Dirac points determine the phases of the
system. Alternately, it can also be understood from band inversion,
which is the mechanism of the TI phase in $HgTe/CdTe$ QWs. The two
ways are equivalent since the occurrence of band inversion
corresponds to changing the sign of one Dirac mass and causes a
topological phase transition, which can't happen without closing the
gap.

Spin-orbit coupling is a necessary condition for the existence of
TI. Its role is to induce a gap in the Dirac dispersion and to
ensure that the gap is finite everywhere in the Brillouin zone (BZ).
It also has been known that non-magnetic and magnetic staggered
potentials can perturb the Dirac dispersion and induce a gap. Such
terms can be obtained by the proximity effect to the corresponding
orders (charge density wave (CDW) and antiferromagnetism (AF))
\cite{hosur1}. When these terms coexist, their interplay will
determine the phase of the system. In this paper, we study the
interplay of these terms. We start from a trivial insulator with
spin-orbit coupling and introduce non-magnetic and magnetic
staggered potentials with check-board and stripe patterns into the
system. We find that when the strength of the potential is strong
enough a band inversion occurs and the system shows QSH effect.
Specially the QSH effect induced by magnetic staggered potential
breaks ${\cal T}$ symmetry, in contrast to the previously studied
QSH effect.

To be concrete, we study a model describing $HgTe/CdTe$ QWs. It
resides on a square lattice with four orbit states $|s,\uparrow
\rangle, |p_x+ip_y, \uparrow \rangle, |s,\downarrow \rangle,
|(p_x-ip_y),\downarrow \rangle$ ($\uparrow, \downarrow$ denote the
electron's spin) on each site. In the momentum space, the
Hamiltonian writes,
\begin{eqnarray}\label{h1}
H_0({\bf k})&=&[4D-2D(\cos k_x+\cos k_y)]I \\ \nonumber
&+&[M+4B-2B(\cos k_x+\cos k_y)] \sigma_z \\ \nonumber
&+&2A \sin k_x s_z \otimes \sigma_x+
2A \sin k_y \sigma_y
\end{eqnarray}
Here$\vec{\sigma}$ and $\vec{s}$ are Pauli matrices representing the
orbits and the electron's spin and $I$ is identity matrix. $A,B,D$
and $M$ are four independent parameters. The tight-binding Hamiltonian can be
directly obtained by a lattice regulation of the effective
low-energy Hamiltonian describing the physics of $HgTe/CdTe$ QWs.
We can also view it as a simple toy model conveniently describing both topological and ordinary
 phases of non-interacting electrons in 2D.
The energy spectrum of $H_0({\bf k})$ has two double degenerate branches
$E_{\bf k}=(4D-D_{\bf k})\pm\sqrt{(2A\sin k_x)^2+(2A\sin k_y)^2+(\tilde{M}-
B_k)^2 }$, where $\tilde{M}=(M+4B)$, $B_{\bf k}=2B(\cos k_x +\cos k_y)$
and $D_{\bf k}=2D(\cos k_x +\cos k_y)$. At half-filling, depending on the
values of $M$ and $B$, the system can be QSH or trivial
insulator.

\begin{figure}[t]
\includegraphics[width=8.4cm]{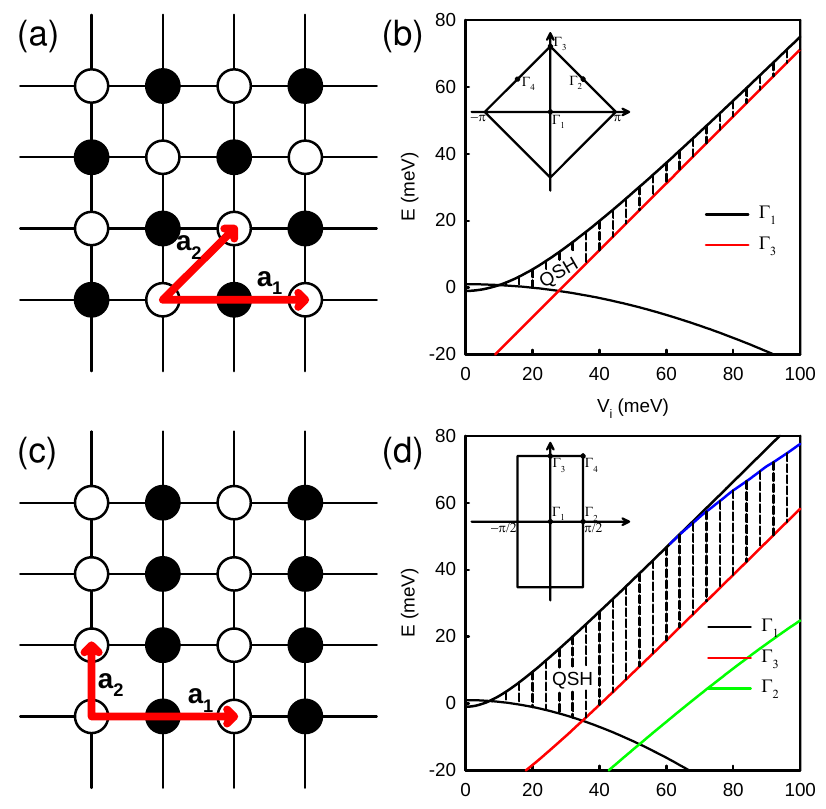}
\caption{(Color online).
The staggered potential with check-board pattern (a) and stripe pattern (c) and
the band evolvements at TRIM ((b) and (d)) with the potential strength corresponding to the patterns in (a) and (c) respectively.
The open and filled circles on lattice sites in (a) and (c) represent on-site potentials with equal values but opposite signs.
The new lattice vectors and new BZs are shown in (a), (c) and the insets of (b), (d).
The blue curve in (d) is a band evolving at other momentum which will determine the gap size at large potential strength.
The range with vertical dashed lines in (b) and (d) marks the gap of the induced topological phase.
The parameters are fixed for all calculations in this paper to be $A=36.45 meV$, $B=27.44 meV$, $D=20.48 meV$ and $M=1 meV$.}
\label{fig1}
\end{figure}

The system defined by Eq. (\ref{h1}) is invariant under ${\cal T}$
and spatial inversion. Since the inversion operator ${\cal P}$
operating on $p-$type orbit generates a minus sign, the inversion
operator writes ${\cal P}=I\otimes \sigma_z$. Thus the signs of
$\tilde{M}- B_{\bf k}$ at the four time-reversal invariant momenta
(TRIM) determine the phase of the system, i.e., $M(M+4B)^2(M+8B)<0
(>0)$ for QSH (trivial) insulator \cite{note1, imura1}. In this
paper, we restrict our calculations in the parameter range
describing $HgTe/CdTe$ QWs, where $B>|M|>0$. Since the TRIM
$\Gamma_1$ (${\bf k}=[0,0]$ in the BZ, see the inset of (b) in Fig.
(\ref{fig1})) dominates the physics in the range, the Hamiltonian
describes QSH effect for $M<0$ and a trivial insulator for $M>0$.
Experimentally the gap parameter $M$ can be continuously tuned from
a positive value for thin QWs with thickness $d<d_c$ to a negative
value for thick QWs with $d>d_d$ ($d_c$ is a critical thickness and
equals $6.3$nm for $HgTe/CdTe$ QWs) \cite{bernevig1, konig1}. The
gap parameter $M$ in Eq. (\ref{h1}) represents an on-site potential,
which has different sign for $s-$type and $p-$type orbits. When
changing its sign, one of the occupied bands changes from $p-$type
($M>0$) to $s-$type ($M<0$). Since the two kind of orbits have
different parities, a band inversion will induce a topological phase
transition.

To drive the system into QSH phase, a band inversion is needed. So it is
interesting to seek ways other than tuning the gap parameter $M$ to
generate the band inversion. Below we fix $M>0$ when the system is a
trivial insulator and find ways to inverse the bands at TRIM
$\Gamma_1$. A natural thought is to enlarge the Hamiltonian, which
make it possible to add more terms to it. Firstly we consider
putting the system on a check-board square lattice (CSL), which can introduce alternating potential with check-board
pattern. In the case, the unit-cell is doubled. The Hamiltonian $H_0({\bf k})$ is
enlarged to $8\times 8$ and becomes,
\begin{eqnarray}\label{h2}
H_1({\bf k})=4DI-2D(\cos k_x+\cos k_y)\tau_x  \\ \nonumber
+(M+4B)\sigma_z-2B(\cos k_x+\cos k_y)\tau_x \otimes \sigma_z \\ \nonumber
+2A \sin k_x \tau_x\otimes s_z \otimes \sigma_x+
2A \sin k_y \tau_x \otimes \sigma_y
\end{eqnarray}
Here $\vec{\tau}$ is Pauli matrix describing the two sublattices. We
have been able to identify two interesting on-site terms: (i) a non-magnetic staggered potential (or CDW potential) $V_1 \tau_z$; (ii) a magnetic
staggered potential (or AF potential) $V_2 \tau_z\otimes s_z$ (The
magnetization can also lie in the plane and couples with the
in-plane components of the electron's spin, generating terms like
$\tau_z\otimes s_x$ or $\tau_z\otimes s_y$. We will discuss these terms later). The former term preserves ${\cal T}$ symmetry
while the latter breaks. Including the above terms to Eq. (\ref{h2}), though
we can't obtain the analytic forms of the energy spectrums, they are
the same for the two different cases. The system remains gapped and the
occupied bands evolve with the strength of $V_i$ ($i=1,2$).

At TRIM $\Gamma_1$, the energy eigenvalues are: (1)
$4D-\tilde{M}-\tilde{B}_{-}$; (2) $4D-\tilde{M}+\tilde{B}_{-}$;
(3)$4D+\tilde{M}-\tilde{B}_{+}$; (4) $4D+\tilde{M}+\tilde{B}_{+}$,
where $\tilde{B}_{\pm}=\sqrt{(4D\pm 4B)^2+V_i^2}$ and each is double
degenerate. The corresponding eigenvectors and their parities can
also be obtained, which are listed in Table I. The band evolvement
with the strength $V_i$ is shown in (b) of Fig. (\ref{fig1}). At
half-filling and for $V_i=0$, bands (1) and (2) are occupied. Since
we choose $M>0$, the system is a trivial insulator. As $V_i$
increases, bands (2) and (3) firstly approach each other and at a
critical value of $V_i$ the filling for the two bands will
interchange. Bands (2) and (3) consist of electrons in the $p-$type
and $s-$type orbits respectively and have opposite parities. For the
case with CDW term the band inversion will induce a topological
phase transition and drive the system into QSH phase. It is
interesting that AF term has the same effect as CDW term. As we
further increase the strength of the potential, the system remains
in QSH phase until another band inversion occurs at other TRIM,
which happens at potential strength bigger than 100 $meV$. The gap
range of the resulting QSH insulator is also denoted in (b) of Fig.
(\ref{fig1}). At small potential strength, it is determined by band
(2) and (3) and increases with increasing potential strength. Then a
band evolving at TRIM $\Gamma_3$ (${\bf k}=[0,\pi]$ in the BZ, see
the inset of (b) in Fig. (\ref{fig1})) becomes the lower restriction
of the gap and restricts its further increase. It is interesting to
note that the behavior is similar to what happens in topological
Anderson insulator, where the phase diagram has to be obtained by
conductivity calculations since the system has no translation
symmetry in the presence of disorder \cite{jli1, hjiang1, groth1,
guo1}.

\begin{table}
\begin{tabular}{|c|c|c|c|}
  \hline
  $No.$ &Eigenvector(CDW)& Eigenvector(AF)& Parity \\
  \hline
  $1$ &$(-\phi_{2-}^2,1^6);(-\phi_{2-}^4,1^8)$&$(-\phi_{2-}^2,1^6);(-\phi_{1-}^4,1^8)$&-1 \\
  \hline
  $2$ &$(\phi_{1-}^2,1^6);(\phi_{1-}^4,1^8)$&$(\phi_{1-}^2,1^6);(\phi_{2-}^4,1^8)$& -1 \\
  \hline
  $3$ &$(\phi_{2+}^3,1^7);(\phi_{2+}^1,1^5)$&$(\phi_{1+}^3,1^7);(\phi_{2+}^1,1^5)$& 1 \\
  \hline
  $4$ &$(-\phi_{1+}^3,1^7);(-\phi_{1+}^1,1^5)$&$(-\phi_{2+}^3,1^7);(-\phi_{1+}^1,1^5)$& 1 \\
  \hline
\end{tabular}
\caption{The eigenvectors for Hamiltonian at TRIM $\Gamma_1$ and
their parities. Here $\phi_{1\pm}=\frac{\tilde{B}_{\pm}+ V_i}{4B\pm
4D}$ and $\phi_{2\pm}=\frac{\tilde{B}_{\pm}- V_i}{4B\pm 4D}$. The
superscripts on the values of the eigenvectors represent the
corresponding position in the eigenvectors.} \label{table}
\end{table}

We can also put the system on a stripe square lattice (SSL), as
shown in (c) of Fig. (\ref{fig1}). The Hamiltonian $H_0({\bf k})$ becomes,
\begin{eqnarray}\label{h3}
H_2({\bf k})=(4D-2D\cos k_y)I-2D \cos k_x \tau_x \\ \nonumber
+(M+4B-2B\cos k_y)\sigma_z-2B\cos k_x \tau_x\otimes \sigma_z+\\ \nonumber
2A \sin k_x \tau_x\otimes s_z \otimes \sigma_x+2A \sin k_y  \sigma_y
\end{eqnarray}
Similarly, a stripe CDW ($\tau_z$) or AF term ($\tau_z\otimes s_z$), which has the same form as its counterpart on
CSL, can be added to the above Hamiltonian. The
energy spectrums for both cases are still the same. The band
evolvements with the potential strength at TRIM are shown in (d) of Fig. (\ref{fig1}).
The CDW or AF term on SSL can also induce a band
inversion and drive the system into QSH phase. However compared to the case in CSL,
the band inversion here occurs at smaller potential strength and the resulting QSH insulator has bigger gap.

So adding CDW or AF term to the system can induce a band
inversion. Though they have different properties under $\cal T$
transformation, they have the same effects on band inversion. The
reason is that the combined Hamiltonian at TRIM $\Gamma_1$ can be decoupled for
each orbit and spin. Each decoupled Hamiltonian has a $2\times 2$
form and has two sub-bands. The CDW or AF term pushes one sub-band
up and the other down, which doesn't depend on the sign of the potential strength.
Since the difference between CDW and AF terms is that AF term acts on
different spin with different sign, they have the same effect on changing the band structure.

The CDW term preserves $\cal{T}$ symmetry and such systems
can be described by $Z_2$ topological invariant \cite{kane2, lfu2}. Since our system
has inversion symmetry, the $Z_2$ invariant can be determined from
the knowledge of the parities of the occupied band eigenstates at the
four TRIM. When the band inversion occurs as we increase the potential strength, the value of the $Z_2$ invariant also change its sign and
becomes non-trivial, indicating that the system is in QSH phase.
For the case with AF term, $\cal{T}$ symmetry is broken and $Z_2$ topological invariant is inapplicable.
However the spin
Hall conductance (SHC) of the resulting QSH phase still shows
quantized value $\frac{e}{2\pi}$ \cite{zgwang1}. So the underlying topological invariant is a spin Chern number, which describes
the quantized spin-Hall conductivity. The concept of spin Chern number has appeared in the recent literature which has its definition for systems with spin $s_z$
conservation \cite{dnsheng1}. For $\cal{T}$ invariant systems, it is equivalent to $Z_2$ topological invariant \cite{lfu2}.

\begin{figure}[t]
\includegraphics[width=8.4cm]{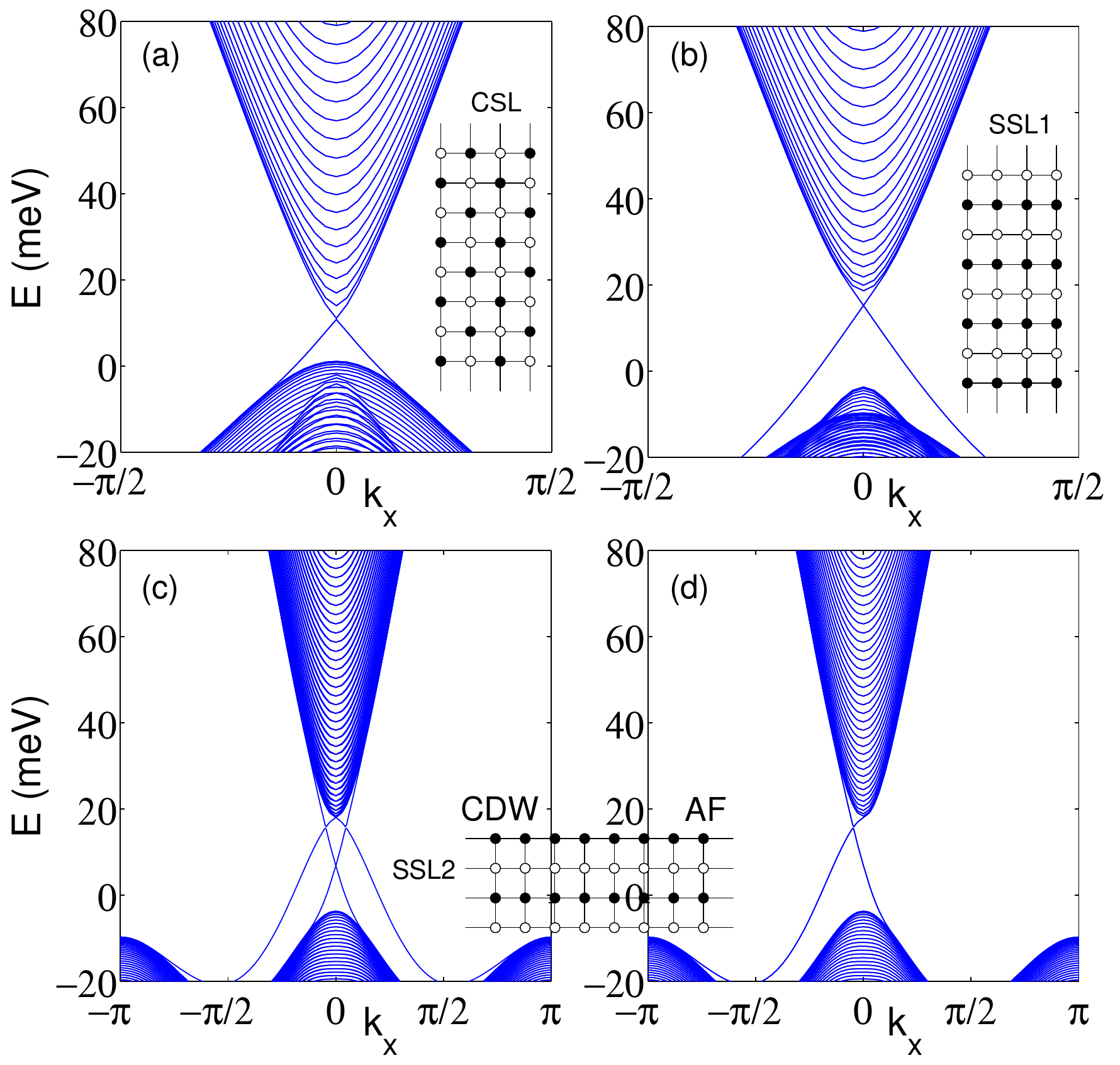}
\caption{(Color online) One-dimensional energy bands with a strip geometry (shown in inset) for (a) check-board pattern and (b), (c), (d) stripe pattern.
In (a) and (b), the energy bands are the same for cases with CDW and AF terms. (c) and (d) have different edges with (b) and the energy bands with CDW term (c)
are different with those with AF term (d). A strip of width $N_y=60$ unit cell with open boundary conditions along $y$ and infinite along $x$ is used
with the staggered potential strength $30 meV$, when the system is in band-inverted phase.}
\label{fig2}
\end{figure}

To further support our identification of the
topological phase, we have performed numerical diagonalization of
Hamiltonian (2) and (3) with CDW or AF term using a strip geometry in the range
of parameters where the system is in band-inverted phase. In accord with the
above arguments we find a pair of spin-filtered gapless states
associated with each edge traversing the gap, which is shown in Fig. (\ref{fig2}). The spin-filtered edge states determine the transport of charge and spin
in the gap range. For a two-terminal device, the conductance is contributed by two conducting channels on the edges and gets quantized value $2e^2/h$.
For a four-terminal device with proper voltage on each terminal, a spin current can be generated \cite{kane1}.

Up to now, we demonstrate the existence of QSH effect induced by non-magnetic and magnetic staggered potential in a trivial insulator with spin-orbit coupling.
In the following, we study the stability of the resulting QSH effect to non-magnetic disorder.
The Hamiltonian we are considering can be decoupled for spin-up and
-down electrons and each describes quantum anomalous Hall (QAH) effect. The resulting QSH effect can be understood as two copies of
QAH effect. For the CDW case, the two copies for spin-up and -down electrons are
related by $\cal{T}$ and the QSH effect is immune to
non-magnetic disorder. However for the AF case, though $\cal T$
symmetry is broken, the system preserves the combined symmetry of
$\cal T$ and a primitive lattice translation, which has been studied in three dimensions and
antiferromagnetic topological insulator is predicted \cite{mong1}. So the two
copies for spin-up and -down electrons in the presence of AF term are related
by the combined transformation. If there is non-magnetic disorder in the
system, the combined symmetry will be broken and the two copies will
behave separately. But we still expect the combined system will be
robust to disorder, because each copy is in QAH phase and
robust to disorder.

\begin{figure}[t]
\includegraphics[width=8.4cm]{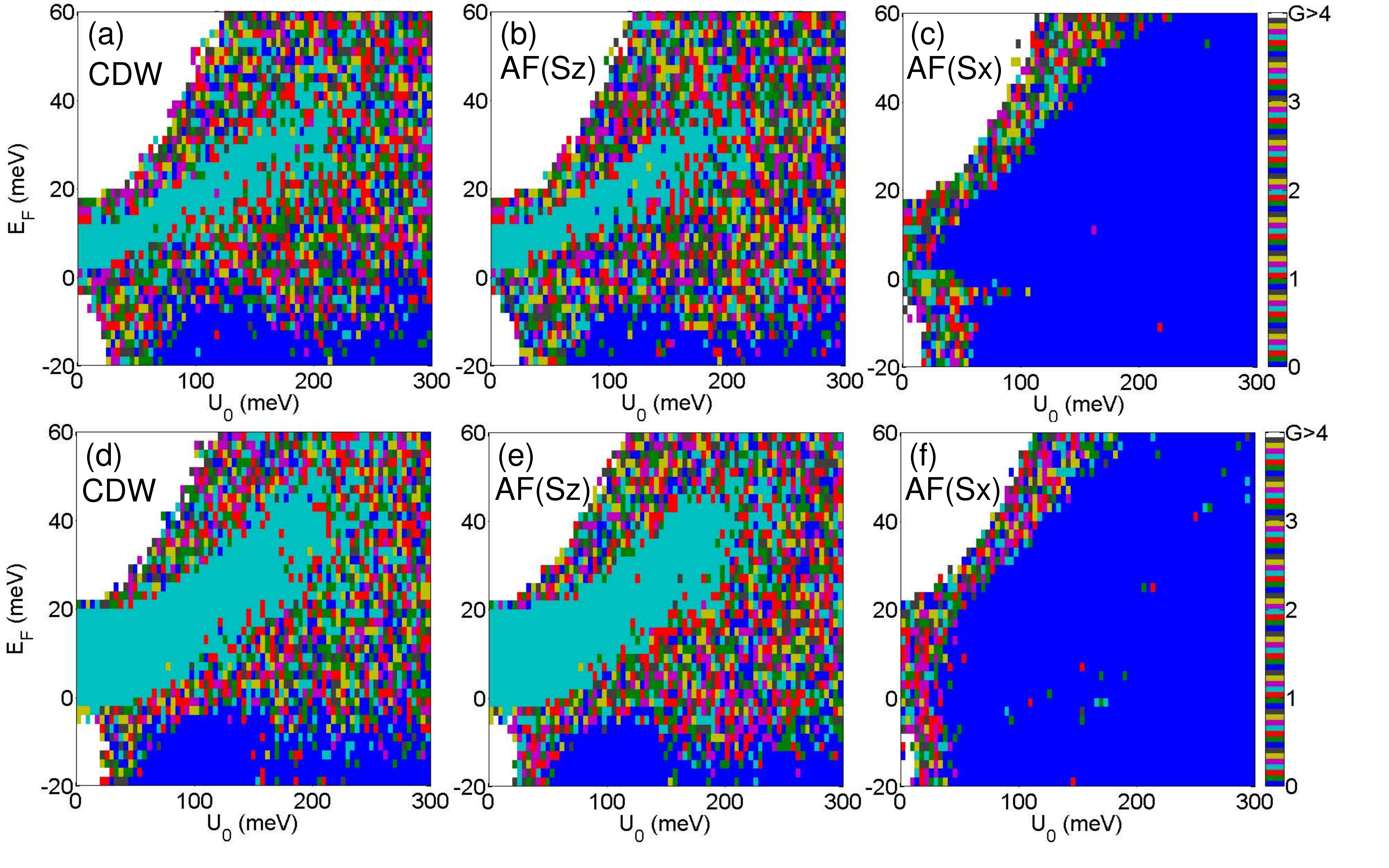}
\caption{(Color online) Conductance $G$ as a function of disorder strength $U_{0}$ and the Fermi level $E_{F}$. Each data point corresponds to
a single disorder realization. (a), (b) and (c) are the results with CDW term, AF terms in $s_z$ and $s_x$ channels on CSL strip;
(d), (e) and (f) are corresponding results on SSL1 strip. The parameters here are the same as those in Fig. (\ref{fig2}), when the system shows band-inverted phase in its clean form.}
\label{fig3}
\end{figure}

To support the above statement, we employ the recursive Green's function method to evaluate the conductance $G$ of
 two-terminal devices (see the insets of Fig. (\ref{fig2})) using Landauer-B\"{u}ttiker formalism. Figure \ref{fig3} shows the results of
such calculations in the space of parameters $(E_F, U_0)$, where $U_0$ is disorder strength and the disorder is described by a random on-site potential uniformly distributed
in the range $(-U_0/2,U_0/2)$. We only consider a single disorder realization at each point of the $(E_F, U_0)$ phase diagram. Nevertheless, this turns
out to be sufficient for studying the stability of the edge states. The reason is that if the edge states are robust to disorder there should be a region
showing quantized conductance $G=2e^2/h$ in the phase diagram and conductance $G$ in the region shows no observable fluctuations, but fluctuates
significantly elsewhere. The plots in Fig. (\ref{fig3}) displays conductance $G$ in a fashion that is designed to amplify the effect of fluctuations.
In (a), (b), (d) and (e) of Fig. (\ref{fig3}), regions showing no observable fluctuations exist in the phase diagrams,
implying that the QSH phase induced
 by CDW or AF (in $s_z$ channel) terms is robust to disorder. We also carried out calculations on strip SSL2
and the results are similar to those on strip SSL1. These results are consistent with what we expect.

\begin{figure}[t]
\includegraphics[width=8.4cm]{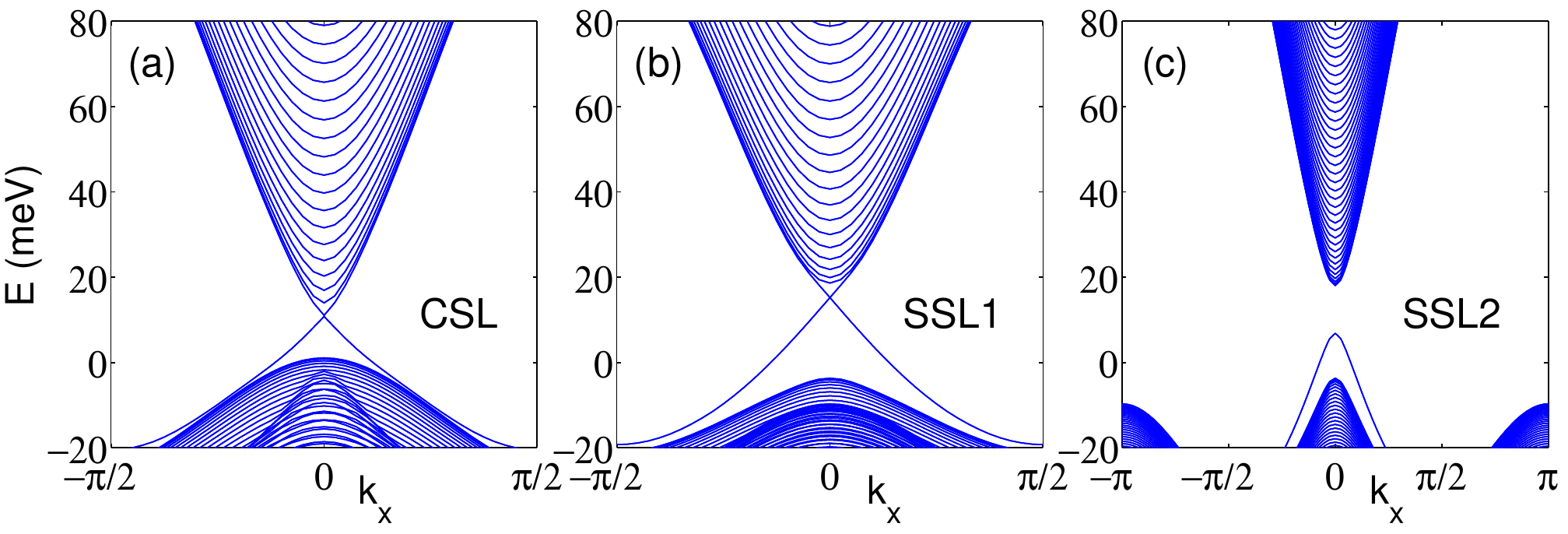}
\caption{(Color online) One-dimensional energy bands with AF term in $s_x$ channel on a
strip geometry (a) CSL; (b) SSL1; (c) SSL2. The parameters used here are the same as those in Fig. (\ref{fig2}). }
\label{fig4}
\end{figure}

As mentioned, for AF term, the magnetization can also lie in the
plane (generating term in $s_x$ or $s_y$ channel) and in the
following we will focus on such term. Including such term to
Hamiltonian described by Eq. (2) or Eq. (3), the band evolvement at
TRIM $\Gamma_1$ is exactly the same as that due to CDW or AF
potential in $s_z$ channel. However unlike the cases with $s_z$
channel term, the combined Hamitonian can't be decoupled for spin-up
and spin-down electrons (i.e. there is no longer $s_z$ conservation
in the system). We perform numerical diagonalization of Hamiltonian
(2) and (3) with AF term in $s_x$ channel (it is the same with $s_y$
channel term) and the one-dimensional energy bands with a strip
geometry are shown in Fig. (\ref{fig4}). For strips CSL and SSL1
(see insets of Fig. (\ref{fig2})), the energy spectrums are similar
to those in Fig. (\ref{fig2}) and there are edge states traversing
the gap in band-inverted phase. But for strip SSL2, the edge states
vanish though the system is in band-inverted phase. The difference
can be understood from the symmetries existing in the system. AF
term in $s_x$ channel couples spin-up and -down electrons and the
system only preserves the combined symmetry of $\cal T$ and a
primitive lattice translation. The existence of edge states is due
to the combined symmetry. Since the edges of strip SSL2 are
ferromagnetic and break the combined symmetry, the edge states are
gapped. However because the combined symmetry on the edges of strips
CSL and SSL1 is still preserved, the edge states exist. But these
edge states are no longer robust to disorder, which can be shown
from conductivity calculations ((c) and (f) of Fig. (\ref{fig3})).
Here it is disorder that breaks the combined symmetry and the QSH
phase in the system will no longer be protected. While for system
with $s_z$ channel term, though ferromagnetic edges or disorder
breaks the combined symmetry, the spin $s_z$ conservation follows to
assure the existence and robustness of the edge states.

In conclusion, we introduce non-magnetic and magnetic staggered potentials to a trivial insulator with spin-orbit coupling and find that they can
induce a topological phase transition and drive the system into topological phase. For non-magnetic staggered potential,
the resulting QSH phase is protected by $\cal T$ symmetry and supports edge state on any edge, which is robust to disorder.
While for magnetic staggered potential, there is a combined symmetry of $\cal T$ and a primitive lattice translation in the system.
If there is also an additional symmetry i.e., spin $s_z$ conservation, edge states will exist on any edge in band-inverted phase and are robust to disorder.
However in the absence of such symmetry, though edge states will exist on specific edges in band-inverted phase,
they are no longer robust to disorder.
Our these results imply that though
generally QSH effect is protected by $\cal T$ symmetry but if there are additional symmetries, QSH effect can also be found in $\cal T$ breaking systems.
.

\emph{Acknowledgment}.--- The authors are indebted to M. Franz,
S.-P. Kou, G. Refael, Q.-F. Sun and J.-W. Ye for stimulating
discussions. Support for this work came from the funds from Beijing
Education Commission under Grant No. KM200910028008, the Ministry of
Science and Technology of China under Grant Nos. 2011CBA00102 and
2011CB921700, NSFC under Grant Nos. 10774015 and 11074023, and the
Research Grant Council of Hong Kong under Grant Nos. HKU 7051/10P
and HKUST3/CRF/09.


\begin{thebibliography}{10}
\bibitem [*] {add} To whom correspondence should be addressed.
\bibitem{note1} The conditions can also be simplified as $\frac{B}{M}<-\frac{1}{8}$ for QSH insulator and
$\frac{B}{M}>-\frac{1}{8}$ for trivial insulator with $M\neq 0$, $B\neq 0$ and $\frac{B}{M}\neq -\frac{1}{4}$.
Here in the range of QSH insulator, $\frac{B}{M}=-\frac{1}{4}$ is also a critical transition point, where the bands close and the system becomes a semi-metal.
Though the syetems at the two sides are still topological non-trivial insulator, the gapless states appear at $k_x=0$ ($k_x=\pi$) for
$\frac{B}{M}>-\frac{1}{4}$ ($\frac{B}{M}<-\frac{1}{4}$).
\end{thebibliography}
\end{document}